# Using AMC and HARQ to Optimize System Capacity and Application Delays in WiMAX Networks

Iwan Adhicandra

**Abstract**—The IEEE 802.16 technology (WiMAX) is a promising technology for providing last-mile connectivity by radio link due to its high speed data rates, low cost of deployment, and large coverage area. However, the maximum number of channels defined in the current system may cause a potential bottleneck and limit the overall system capacity. The aim of this paper is to compare the impact on system performance of different solutions used to mitigate the impairments due to the radio channel. In particular, taking into account the WiMAX system capacity as well as application delays, the paper presents the simulation results obtained when a static QPSK ½ Modulation and Coding Scheme (MCS) is adopted. Then, the study is aimed at evaluating the improvements introduced by the adoption of an adaptive modulation and coding (AMC) and an AMC jointly with Hybrid Automatic Repeat reQuest (HARQ). Results indicate that the best strategy is to use an aggressive AMC table with the HARQ.

———————————— ◆ ————————————

## 1 INTRODUCTION

THE IEEE 802.16 technology (WiMAX) is a promising alternative to 3G or wireless LAN for providing last-mile connectivity by radio link due to its large coverage area, low cost of deployment and high speed data rates. The standard specifies the air-interface between a Subscriber Station (SS) and a Base Station (BS). The IEEE 802.16-2004 standard [1], also known as 802.16d, was published in October 2004. This was further developed into the mobile WiMAX standard referred to as IEEE 802.16e-2005 or 802.16e [2] to support mobile users. The mobile WiMAX air interface adopts Scalable Orthogonal Frequency Division Multiple Access (SOFDMA) for improved multipath performance in non line of sight (NLOS) environments.

In WiMAX system, there are a number of modulation and forward error correction (FEC) coding schemes, which can be supported. Based on channel conditions, the scheme can be changed on a per frame and per user basis. In order to maximize throughput in a channel varying in time, one of the effective mechanism, namely Adaptive and Modulation Coding (AMC) can be used. Typically, the algorithm calls for the use of the highest modulation and coding scheme that can be supported by the signal-to-noise and interference ratio at the receiver. In this way, the highest possible data rate that can be supported in their respective links is provided to each user. However, it is likely that data transmitted through the air is corrupted because of background noise, channel interferences, etc. In this case, Hybrid Automatic Repeat reQuest (HARQ) can be used to provide a reliable way and to ensure that packets can be successfully received in order. In order to designing the system, the HARQ can be used to reduce system bit error rate (BER). By combining flexible channelization with the AMC, it enables mobile WiMAX technology to improve both system capacity and coverage.

There have been some studies examining the performance of WiMAX Networks using the AMC and the HARQ. Lynn et al [3] evaluated the performance of asymmetric Time Division Duplex (TDD) system that employs the AMC and the HARQ, with consideration of the effect of control delays in TDD. Moh et al [4] proposed an enhanced HARQ, which would determine the number of multiple copies needed based on the channel feedback. Moreover, Saeed et al [5] studied Dynamic HARQ (DHARQ) for WiMAX. They proposed a power control scheme for WiMAX multihop relay system. Although much work has been done to date [8-14], more studies need to be conducted to ascertain the effects of the AMC and the HARQ in optimizing network capacity and application delays.

The purpose of this study was to design a WiMAX system configuration to optimize both network capacity and application delays since the capacity and the application delays usually have a trade-off. We used the AMC and the HARQ to tune the WiMAX network such that its adaptations improve both performance capacity and delay. Firstly, we examined the performance of the WiMAX network with respect to capacity usage and application delays using static Quadrature Phase Shift Keying (QPSK) ½ Modulation and Coding Scheme (MCS).

————————————————

• *I. Adhicandra is with the Department of Information Engineering, The University of Pisa, Via G. Caruso 16, 56122 Pisa, Italy.*

Secondly, we used the AMC to compare performance with that of QPSK ½ setting. Then, we used an aggressive AMC table to study its impact on capacity usage and application delays. Finally, we used AMC table in conjunction with the HARQ.

This paper is organized as follows. Section 2 explains an overview of the AMC and the HARQ. Section 3 provides the design of system models. Results and discussions are presented in Section 4. Finally, Section 5 concludes the paper.

## 2 BACKGROUND

In order to take advantage of fluctuations in the channel, WiMAX systems can use adaptive modulation and coding. The principles are as following: in order to avoid excessive dropped packets, it transmits at a lower rate when the channel is poor and it transmits as high a data rate as possible when the channel is good. To achieve lower data rates, a small constellation can be used, such as QPSK, and low-rate error-correcting codes, such as turbo or rate convolutional codes. To achieve the higher data rates, large constellations can be used, such as 64 QAM, and less robust error correcting codes, for example, turbo, rate convolutional, or low-density parity check (LDPC) codes. In total, there are 52 possible configurations of modulation order, coding types and rates, although only a fraction of these can be offered in most implementations of WiMAX.

Figure 1 shows a block diagram of an AMC system. Firstly, we consider a single-user system attempting to transmit as fast as possible through a channel with a variable signal-to-interference-plus-noise ratio (SINR), for example, due to fading. Then, the transmitter needs to transmit data from its queue as fast as possible, subject to the data being demodulated and decoded reliably at the receiver. For adaptive modulation and coding, feedback is critical. The transmitter needs to know the "channel SINR" $\gamma$, which can be defined as the received SINR $\gamma_r$ divided by the transmit power $P_t$, which itself is usually a function of $\gamma$. Thus, the received SINR is $\gamma_r = P_t \gamma$.

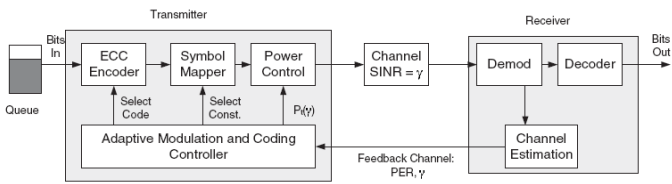

Fig. 1. Adaptive modulation and coding block diagram.

In the AMC, the main objective is to efficiently control three quantities at once:, coding rate, transmit rate (constellation), and transmit power. This corresponds to developing an appropriate policy for the AMC controller. In practice, developing and fine-tuning the algorithm, based on extensive simulations, need to be done by the system engineer since performance depends on many factors, although reasonable guidelines can be developed from a theoretical study of adaptive modulation. These considerations include block error rate (BLER) and received SINR, Automatic repeat request (ARQ), Power control versus water filling, and Adaptive modulation in OFDMA.

On the other hand, the ARQ supports rapid retransmissions, and hybrid-ARQ generally can increase the ideal BLER operating point by about a factor of 10 for example, from 1 percent to 10 percent. It may be possible to accept a BLER approaching even 70 percent for delay tolerant applications.

## 3 SYSTEM MODELS

In this experiment, we used OPNET Modeler version 14.5 with WiMAX Module capability [6]. We designed four scenarios including "Using QPSK ½ as the Static MCS", "Using Adaptive Modulation and Coding (AMC)", "Using Conservative AMC", and "Using HARQ with AMC A". The network was composed of one WiMAX cell (i.e. a single Base Station, BS) and 20 Subscriber Station (SS) nodes (Figure 2). The parameters of SSs and BS can be seen at Table I and Table II. Each SS represents a TCP client executing an upload towards a TCP server located in the wired part of the network scenario. The channel was configured to vary according to ITU Pedestrian A multipath fading model. For traffic configuration, all SS nodes had an uplink application load of 20 Kbps for a total of 0.4 Mbps. All SS nodes were configured to use QPSK ½ for the uplink application. The cell used an SOFDMA frame with 512 subcarriers and a frame duration of 5 milliseconds. For QPSK ½, the expected uplink

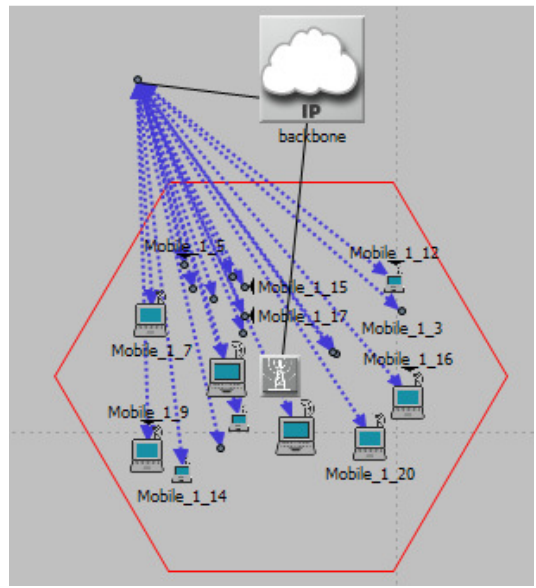

Fig. 2. Network Model.

capacity was around 0.59 Mbps.

TABLE 1
SS PARAMETERS

| Antenna Gain (dBi) | -1 dBi |
|---|---|
| Max Transmission Power (W) | 0.05 |
| PHY Profile | WirelessOFDMA 20 MHz |
| PHY Profile Type | OFDM |
| Type of SAP | IP |
| Modulation and Coding | QPSK1/2 |
| Average SDU Size (bytes) | 1500 |
| Activity Idle Timer (second) | 60 |
| Buffer Size (bytes) | 64 KB |
| Pathloss Model | Free Space |
| Terrain Type (suburban Fixed) | Terrain Type A |
| Ranging Power Step (mW) | 0.25 |
| Piggyback BW Request | Enabled |
| CQICH Period | 3 |
| Contention-Based Reservation Timeout | 16 |
| Request Retries | 16 |

TABLE 2
BS PARAMETERS

| Antenna Gain (dBi) | 15 dBi |
|---|---|
| Max Transmission Power (W) | 0.5 |
| Minimum Power Density (dBm/subchannel) | -110 |
| Maximum Power Density (dBm/subchannel) | -60 |
| Ranging Backoff Start | 2 |
| Ranging Backoff End | 4 |
| Bandwidth Request Backoff Start | 2 |
| Bandwidth Request Backoff End | 4 |
| PHY Profile | WirelessOFDMA 20 MHz |
| PHY Profile Type | OFDM |
| Type of SAP | IP |

TABLE 3
AMC A

|   | Mandatory Exit Threshold (dB) | Minimum Entry Threshold (dB) | Modulation and Coding |
|---|---|---|---|
| 0 | -20 | 2.0 | QPSK ½ |
| 1 | 5.0 | 5.9 | QPSK ¾ |
| 2 | 8.0 | 8.9 | 16-QAM ½ |
| 3 | 11 | 11.9 | 16-QAM ¾ |
| 4 | 14 | 14.9 | 64-QAM ½ |
| 5 | 17 | 17.9 | 64-QAM 2/3 |
| 6 | 19 | 19.9 | 64-QAM ¾ |

TABLE 4
AMC B

|   | Mandatory Exit Threshold (dB) | Minimum Entry Threshold (dB) | Modulation and Coding |
|---|---|---|---|
| 0 | -20 | 2.0 | QPSK ½ |
| 1 | 11 | 11.9 | QPSK ¾ |
| 2 | 14 | 14.9 | 16-QAM ½ |
| 3 | 17 | 17.9 | 16-QAM ¾ |
| 4 | 20 | 20.9 | 64-QAM ½ |
| 5 | 23 | 23.9 | 64-QAM 2/3 |
| 6 | 25 | 25.9 | 64-QAM ¾ |

**A. Scenario 1: Using QPSK ½ as the Static MCS**

In this scenario, we implemented QPSK ½ modulation and coding scheme. QPSK ½ is a conservative MCS with bits to symbol ratio of 1. Using QPSK ½ may reduce the block error rate, and we estimated TCP retransmissions were lower as well improving application performance. On the other hand, it used more symbols to transmit the same data causing consumption of system radio resources.

**B. Scenario 2: Using Adaptive Modulation and Coding (AMC)**

In this scenario, we turned on AMC functionality on the service flows of the SS nodes and chose an AMC table for the BS to use on the uplink (Table III). Using the AMC is advantageous in the sense that it automatically adjusts to an appropriate MCS given an SNR value. If the SNR is good, a node uses a MCS more efficient than the basic QPSK ½. We determined if using the AMC had any disadvantages at all against static MCS of QPSK ½.

**C. Scenario 3: Using Conservative AMC**

In this scenario, we switched to conservative MCS at higher SNR values (Table IV). The aim was to reduce BLER, but might use more system capacity. We examined the physical layer block error rate and compared with QPSK ½ scenario.

**D. Scenario 4: Using HARQ with AMC A**

In this scenario, we used a conservative AMC table to improve application performance significantly at the expense of system capacity. Since we do not want to lose the advantages of AMC A (less system capacity usage), but at the same time, we do not want to cause TCP retransmissions that significantly degrade application performance, so that we decided to keep AMC table A, but used the HARQ on service flows. Using the HARQ can give dual advantages of SNR gain and fast retransmissions. We configured the HARQ on the SS nodes.

## 4 RESULTS

All figures in this section use time duration (in seconds) for the x-line and use the values indicated in the legend (eg. bits/sec, %, etc) for the y-line.

In the scenario 1, both the throughput and the load were around 480 Kbps (Figure 3). This was equal to the application load plus the TCP and IP headers. We saw that WiMAX delay was high and UL frame usage was around 81% showing that QPSK ½ MCS was consuming large system resources (Figure 4 and Figure 5). Thus application delays have a trade off with system capacity.

In the scenario 2, we noticed that WiMAX delays were less (Figure 7). We saw that UL frame usage was around 40% only compared to 81% with QPSK ½ (Figure 8). Thus the AMC did gain in system capacity. However, we found that AMC table A was too aggressive for both the throughput and the load (Figure 6). To solve this, the strategy was to use a conservative AMC table which can reduce BLER, but might use more system resources.

In the scenario 3, we examined the physical layer block error rate and found that it has increased compared to the

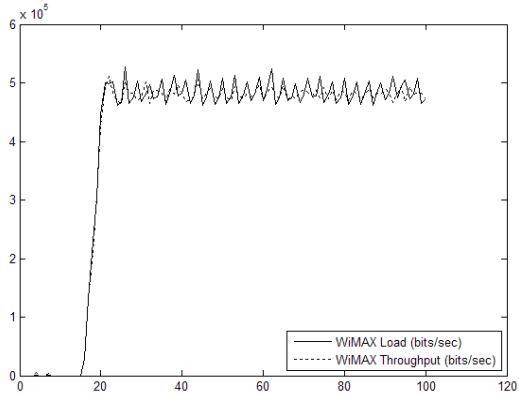
Fig. 3. WiMAX Performance with QPSK.

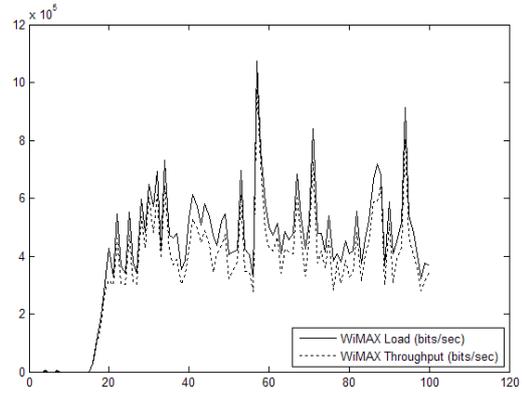
Fig. 6. WiMAX Performance with AMC A.

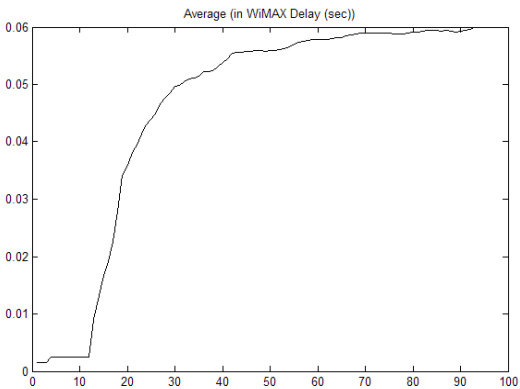
Fig. 4. Average WiMAX Delay with QPSK.

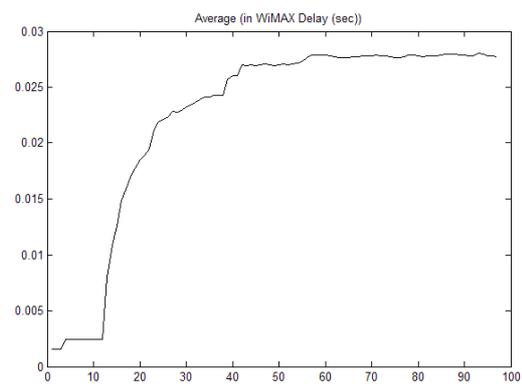
Fig. 7. Average WiMAX Delay with AMC A.

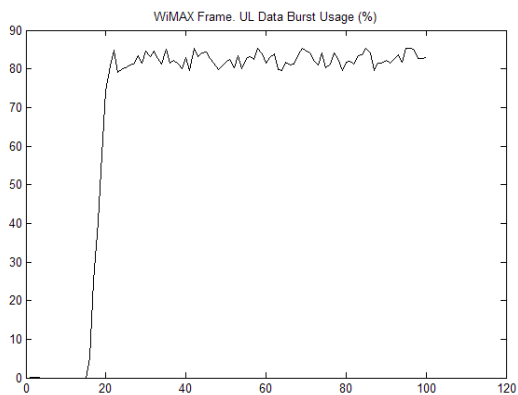
Fig. 5. WiMAX Frame UL Data Burst Usage with QPSK.

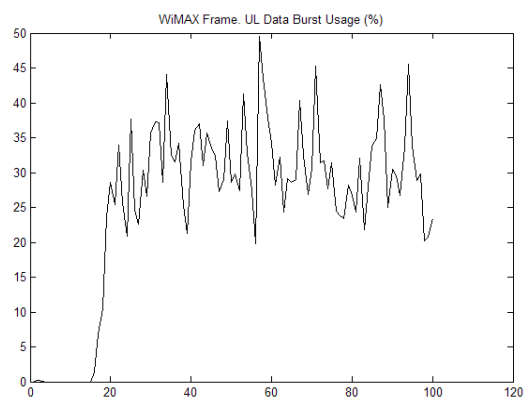
Fig. 8. Frame UL Data Burst Usage with AMC A.

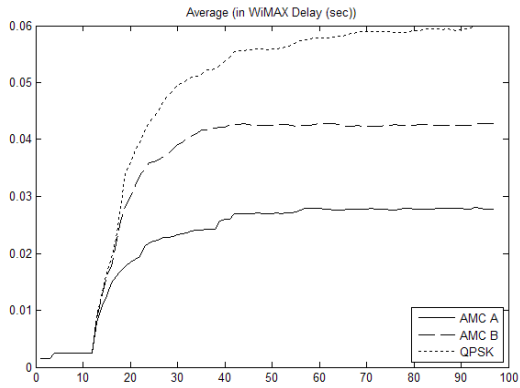

Fig. 9. Average WiMAX Delay with different AMC.

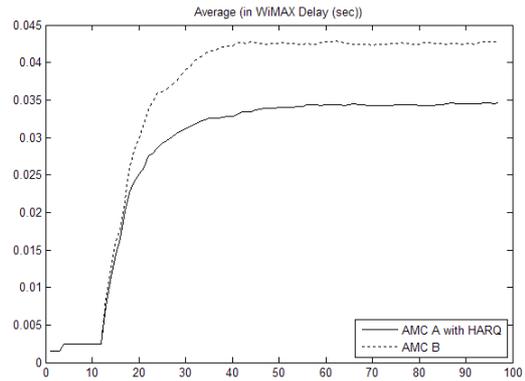

Fig. 12. WiMAX Delay with different AMC and HARQ.

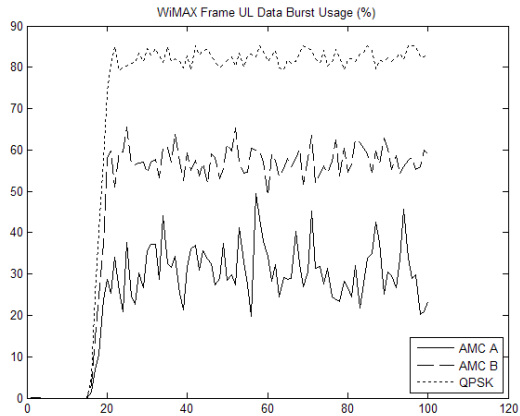

Fig. 10. WiMAX Frame UL Data Burst Usage with different AMC.

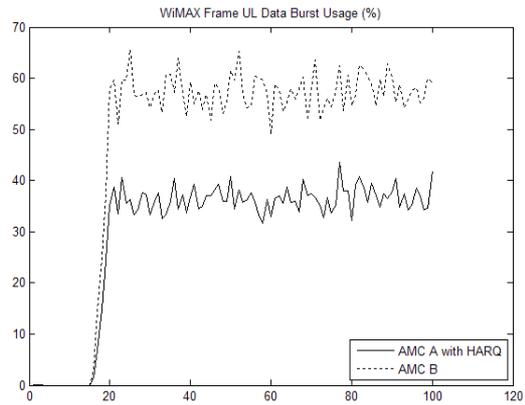

Fig. 13. WiMAX Frame UL Data Burst Usage with different AMC and HARQ.

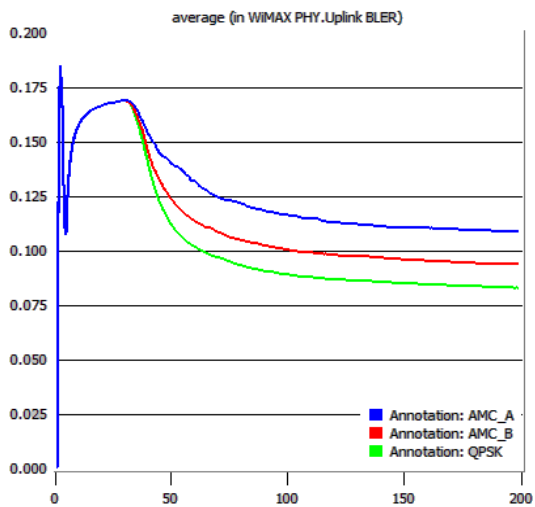

Fig. 11. Average WiMAX UL BLER with different AMC and QPSK.

scenario 1. Since more packets were dropped at the physical layer, retransmissions ensued degrading application performance. We compared the performance of QPSK, different AMC table, and BLER. WiMAX delays for AMC B were less than that for QPSK ½ but more than that for AMC A, since average symbols used in AMC A were lesser (Figure 9). We saw that using AMC B caused more system resources to be used (Figure 10). Thus by increasing the system resources usage, we improved application performance. The BLER for AMC B has reduced from AMC A, which was the reason application performance has improved (Figure 11).

In the scenario 4, we saw that WiMAX delays for this scenario had reduced, since we switched back to AMC A and were using the same number of average symbols (Figure 12). Finally we observed that switching back to AMC A and using the HARQ indeed was the optimal strategy. Radio resource usage had reduced back to 40%, while TCP delays were also small (Figure 13). Both capacity and application delay criteria have now been optimized.

## 7 CONCLUSION

In this paper, we have conducted a study to find optimal configuration of the WiMAX network with respect to system capacity and application delays. By using QPSK½, delays were very low showing that one of the optimization criteria which were application delays was satisfied, and QPSK ½ MCS was consuming large system resources. By using AMC table A, average delays had increased to very high values, and we found that AMC table A was too aggressive. By using conservative AMC, the BLER for AMC B has reduced from AMC A, which was the reason performance has improved. By using the HARQ with AMC A, radio resources usage had reduced back while average delay were also small, so that both capacity and application delay criteria have now been optimized. We conclude that the best strategy is to use an aggressive AMC table with the HARQ.

**Iwan Adhicandra** completed his graduate and postgraduate studies in electrical engineering from Trisakti University, Jakarta, Indonesia, and the University of Sheffield, United Kingdom in 1998 and 2001 respectively. He also completed his postgraduate studies in electrical engineering from Politecnico di Torino, Italy in 2007. From 2000 to 2007 he had been working as a researcher in the Centre for Mobile Communications Research (C4MCR), the University of Sheffield, United Kingdom, in the Computer Communications Research Group, Leeds Metropolitan University, United Kingdom, and in the Connectivity Systems and Networks (COSINE) Group, Philips Research Europe, Eindhoven, the Netherlands. Since October 2007 he has been working with the Department of Information Engineering at the University of Pisa, Italy. He is a member of the IEEE, the ACM, the IET, and the BCS.